\documentclass{article}
\usepackage{spconf,amsmath,graphicx}
\usepackage{amssymb}

\usepackage{booktabs}
\usepackage[hyphens]{url}
\usepackage{caption} 
\usepackage{hyperref}
\usepackage[style=ieee]{biblatex}
\usepackage{paralist}
\usepackage[disable]{todonotes} % Add [disable] for the submission
\usepackage{comment}

\DeclareMathOperator*{\argmax}{argmax}

\title{Improving vision-inspired keyword spotting using dynamic module skipping in streaming conformer encoder}

\name{Alexandre Bittar, Paul Dixon, Mohammad Samragh, Kumari Nishu, Devang Naik}
\address{Apple}

\begin{document}
\sloppy
\maketitle
%
%%%% 0.) ABSTRACT %%%%
\begin{abstract}
    Using a vision-inspired keyword spotting framework, we propose an architecture with input-dependent dynamic depth capable of processing streaming audio. Specifically, we extend a conformer encoder with trainable binary gates that allow us to dynamically skip network modules according to the input audio. Our approach improves detection and localization accuracy on continuous speech using Librispeech top-1000 most frequent words while maintaining a small memory footprint. The inclusion of gates also reduces the average amount of processing without affecting the overall performance. These benefits are shown to be even more pronounced using the Google speech commands dataset placed over background noise where up to 97$\%$ of the processing is skipped on non-speech inputs, therefore making our method particularly interesting for an always-on keyword spotter.
\end{abstract}

%%%% 1.) INTRO %%%%
\section{Introduction}
\label{sec:intro}

Recent advances in deep learning have given rise to numerous end-to-end automatic speech recognition (ASR) models \cite{Radford2023, Zhang2023, Zhang2022, Xu2021, Chung2021, Chan2021, Li2021, Zhang2020} typically trained with encoder-decoder architectures on vast amounts of data. Although capable of approaching human-like ASR performance, such models usually involve substantial memory and power requirements to be stored or to process long audio streams of data directly on portable devices. More efficient approaches using smaller models with limited vocabulary can alternatively be used to solve the simpler task of accurately spotting a set of selected keywords or key-phrases in streaming audio. Such detections can then trigger additional processing from larger ASR-like models, so that the heavy computations only happen on desired audio portions. This paper focuses on improving both the performance and the efficiency of keyword spotting (KWS) models by borrowing techniques from ASR and computer vision.

Similarities between keyword spotting and object localization have already led to a variety of vision-inspired KWS approaches \cite{Segal2019, Wei2021, Zhao2022, Samragh2023}. Indeed, an audio segment can be treated as a 1D-image making computer vision methods applicable. The task of object localization is typically solved using bounding boxes \cite{Liu2016, Redmon2018, Zhou2019}. Precisely localizing keywords can be crucial both for privacy and efficiency considerations, as the ASR model should only be triggered by the lighter KWS model when desired. The base framework for this paper is that of recent work by Samragh \textit{et al.} \cite{Samragh2023}. From the encoded speech representations of a fully convolutional BC-ResNet \cite{Kim2021} backbone, their KWS model yields three types of keyword predictions, namely detection, classification and localization. Using the same general pipeline, we focus on improving the encoder by taking inspiration from ASR models while retaining streaming and low-memory constraints.

The conformer \cite{Gulati2020} architecture, which combines convolution and attention based modules, constitutes a state-of-the-art choice of model for ASR. On top of its proven representational capabilities for speech, the conformer also uses residual connections in a way that allows the inclusion of gates to dynamically skip modules. Recent work by Peng \textit{et al.} \cite{Peng2023} has shown that binary gates can be added inside a Transformer-based ASR architecture to dynamically adjust the network's depth and reduce the average number of computations while retaining the same word error-rate. We extend their input-dependent dynamic depth (I3D) method to the conformer by placing local gates to skip feedforward, convolution and attention modules based on characteristics of the input to the module itself. Skipping modules still requires the full model to be loaded and does not noticeably impact the user-perceived latency, nevertheless, it improves the efficiency and can lead to considerable power savings.

In this paper, we replace the BC-ResNet encoder used in the vision-inspired KWS framework of \cite{Samragh2023} with an I3D-gated conformer that can process streaming audio. Considering the following goals, we aim to:
\begin{enumerate}
    \itemsep0em 
    \item Improve keyword detection and localization while reducing the memory footprint using the conformer.
    \item Minimize computations and lower the power consumption at inference using the I3D method.
\end{enumerate}
We apply this method to (i) Librispeech \cite{librispeech} top-1000 most frequent words, and to (ii) the Google speech commands (GSC) \cite{Warden2018} with added background noise. The former represents the task of detecting occurrences of specific words in continuous speech and is used to assess the encoder's detection and localization capabilities. The latter simulates an audio stream of background noise over which short speech commands are placed in isolation. This allows us to showcase the efficiency benefits of using I3D gates as we expect the model to be able to skip even more processing on non-speech audio regions. Our experiments show that the proposed gating mechanism can skip on average 30\%, 42\%, and 97\% of the encoder modules when processing continuous speech, isolated keywords and background noise respectively.

%%%% METHODS %%%%
\section{Methods}

\begin{figure*}[ht]
\centering
\includegraphics[width=0.89\textwidth]{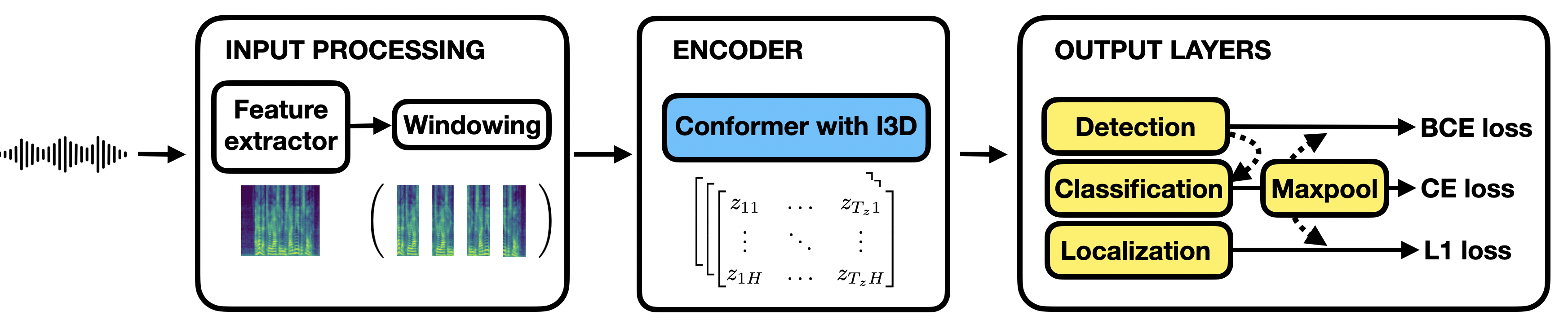}
\caption{Keyword spotting pipeline in training mode.}
\label{fig:pipeline} 
\end{figure*}

The overall KWS pipeline is illustrated in Figure \ref{fig:pipeline}. Compared to the BC-ResNet encoder used in \cite{Samragh2023}, the conformer combines the power of self-attention to also learn global interactions along with convolutions which capture local correlations. The proposed method operates in streaming mode using windowing and handles variable command lengths using max-pooling.

\subsection{Input processing}

The input audio is processed to generate 40-channel Mel filter banks from 25ms windows shifted by a stride of 10ms. To accomodate streaming conditions and limit the attention context, the inputs to the encoder are configured as 1.2 second windows (120 frames) with a shift of 240ms (24 frames). The encoder therefore receives inputs $x\in\mathbb{R}^{B\times 120\times 40}$, where $B$ is the batch size (set to one for inference) and 120 and 40 represent the number of frames and Mel features respectively. During training, these sliding windows are computed at once from the full input utterance and stacked over the batch dimension. At inference time, the model stores the last 960ms of the current window, waits for 240ms, then combines the two together to form the next 1.2 second window and repeats the procedure, which enables it to process streaming audio.

\subsection{Encoder}
The encoder is defined as a standard conformer \cite{Gulati2020} with additional I3D gates \cite{Peng2023} that allow its modules to be dynamically skipped. The input first goes through a subsampling convolutional layer which reduces its time dimension from 120 to $T_z=29$. Its feature dimension also gets projected to the desired hidden size $H$. It then passes through a series of conformer blocks, where each block represents a sequence of (i) feedforward, (ii) mutli-headed self attention, (iii) convolution and (iv) feedforward modules. It is worth noting that none of these modules alter the shape of the transmitted tensors. A local binary gate can therefore be added to the residual connection of each of these four modules in all $N_z$ conformer blocks, so that an input $x\in\mathbb{R}^{B\times T_z\times H}$ gets mapped as,
\begin{equation}
    x\rightarrow x+g(\theta, x)\cdot\text{module}(x) \, ,
\end{equation}
where $g(\theta, x)\in\{0,1\}^B$ is the gating function parameterized by $\theta$. In our experiments, we use a linear layer with weights $W_g \in \mathbb{R}^{H \times 2}$ and bias $b_g \in \mathbb{R}^2$ to implement each gate.
The underlying probabilities of keeping or skipping the related module $\boldsymbol{p_{g}}(x)=\big(p_{\text{keep}}, p_{\text{skip}}\big) \in[0,1]^{B\times 2}$  are then simply computed as,
\begin{equation}
    \boldsymbol{p_g}(x)=\text{Softmax}\big[W_g\cdot \bar{x} + b_g\big] \, ,
\end{equation}
where $\bar{x}$ represents the mean of $x$ taken over the time dimension. The softmax ensures that ${p_{\text{keep}}=1-p_{\text{skip}}}$. During training, the Gumbel-Softmax trick \cite{Jang2017, Maddison2017} is used to sample discrete zeros or ones from $p_{\text{keep}}(x)$ in a differentiable way and obtain the desired binary values for $g(\theta, x)$. At inference time, $g(\theta, x)$ is alternatively computed as $p_{\text{keep}}(x) > \beta$ using a fixed threshold $\beta=0.5$. A regularizer $\mathcal{L_{\text{gate}}}=\lambda\cdot f_{\text{open}}$ with hyperparameter $\lambda=1$ is also defined during training to minimize the fraction of open gates $f_{\text{open}}$ given by,
\begin{equation}
    f_{\text{open}}=\frac{1}{4\, N_z\, B}\sum_{b=1}^{B}\sum_{l=1}^{N_z}\sum_{m=1}^{4}g(\theta_{l,m}, x_b) \, ,
\end{equation}
where $g(\theta_{l,m}, x_b)\in\{0,1\}$ corresponds to the gate output of the $m$-th module in the $l$-th conformer block, when applied to the $b$-th batch element of $x$. We obtained better results by first pretraining the network without gates during a few epochs before enabling them. Our method is also applicable to fine-tune gates on top of a non-gated pretrained conformer.

\subsection{Output layers}

%%%% Detection %%%%
\textbf{Detection.} 
After the encoder, a feedforward layer with a sigmoid activation maps the encodings $z\in\mathbb{R}^{B\times T_z\times H}$ to keyword detection probabilities $\hat{y}_{\text{det}}\in[0,1]^{B\times T_z\times C}$ as
\begin{equation}
\hat{y}_{\text{det}}=\text{Sigmoid}\big[W_{\text{det}}\,z+b_{\text{det}}\big] \, ,
\end{equation}
where $W_{\text{det}}\in\mathbb{R}^{H\times C}$ and $b_{\text{det}}\in\mathbb{R}^{C}$ for $C$ keyword  classes. \\[+0.5ex]
%%%% Classification %%%%
\noindent\textbf{Classification.} 
For the classification probabilities, a feedforward layer is combined with a binary mask to discard all classes with detection probabilities below 0.5, and a softmax activation outputs the final classification probabilities $\hat{y}_{\text{class}}$,
\begin{equation}
    \hat{y}_{\text{class}}=\text{Softmax}\Big[\big(\hat{y}_{\text{det}}\geq0.5\big)\cdot\big(W_{\text{class}}\,z+b_{\text{class}}\big)\Big] \, .
\end{equation}
Here an (unmasked) additional class with label $C+1$ is used to account for situations where no known keyword is present, so $W_{\text{class}}\in\mathbb{R}^{H\times C+1}$ and $b_{\text{class}}\in\mathbb{R}^{C+1}$. \\[+0.5ex]
%%%% Localization %%%%
\noindent\textbf{Localization.} 
A feedforward layer with $W_{\text{loc}}\in\mathbb{R}^{H\times 2C}$ and $b_{\text{loc}}\in\mathbb{R}^{2C}$ simply predicts keyword widths and offsets as
\begin{equation}
(\hat{y}_{\text{width}},\hat{y}_{\text{offset}})=W_{\text{loc}}\,z+b_{\text{loc}} \, .
\end{equation}
%%%% Maxpool %%%%
\noindent\textbf{Maxpool.} 
To account for the variability of keyword lengths, a max-pooling layer is applied over the $T_z=29$ time steps of $\hat{y}_{\text{class}}$ with a kernel-size of 24 and a stride of 1, representing 1 second and 40ms respectively. The selected indices are then used to index the other outputs $\hat{y}_{\text{det}}$, $\hat{y}_{\text{width}}$ and $\hat{y}_{\text{offset}}$, so that all return six time steps per 1.2 second window. In a similar fashion to \cite{Wang2023}, max-pooling allows the loss to only optimize steps with highest posterior probabilities.

\subsection{Ground truths}

During training, after going through the model, the $N_w$ sliding input windows are transferred from the batch axis back to the time dimension, resulting in $T=6N_w$ output steps for the complete utterance. For the ground truths, we therefore consider a receptive field of $R=1$ second with a stride of $S=40$ms so that it matches the model's predictions. We briefly explain the treatment of labels and losses here but refer the reader to~\cite{Samragh2023} for more details. \\[+0.5ex]
\noindent\textbf{Detection.} The event detection labels are computed with the intersection over ground truth (IOG) overlap metric. For a keyword $c$ with begin and end timings $(b,e)$ it is given as
\begin{equation}
    \text{iog}_{t,c}=\frac{\text{overlap}\big[(tS,tS+R), (b,e)\big]}{e-b} \, ,
\end{equation}
for $t=1,\hdots,T$. The detection labels are then computed as,
\begin{equation}
    y_{\text{det}}^{t,c}=\begin{cases}
        1, &\text{if iog}_{t,c}>0.95 \\
        0, &\text{if iog}_{t,c}<0.5 \\
        \text{undefined}, &\text{otherwise} \, .
    \end{cases}
\end{equation}
When the overlap is in between 0.5 and 0.95, it is not clear whether the keyword should be counted as present or absent. Such situations are therefore masked so that no gradient update takes place, resulting in a thresholded binary cross-entropy (BCE) loss. \\[+0.5ex]
\noindent\textbf{Classification.} In order to make the model more robust to keyword collision and confusion, the softmax classifier is trained with a thresholded cross-entropy (CE) loss, where the labels $y_{\text{class}}^{t}$ are defined as
\begin{equation}
    y_{\text{class}}^{t}=\begin{cases}
        k, &\text{if } \exists c\leq C\text{ with iog}_{t,c}>0.95 \\
        C+1, &\text{if iog}_{t,c}<0.05\quad\forall c\leq C \\
        \text{undefined}, &\text{otherwise} \, ,
    \end{cases}
\end{equation}
and $k=\operatorname*{argmax}_c \text{iog}_{t,c}$. \\[+0.5ex]
\noindent\textbf{Localization.} For the keyword localization, the CenterNet approach from \cite{Zhou2019} is adopted. The receptive field center at timestep $t$ is defined as $c_t=t+\frac{R}{2S}$, so that the ground truth width and offset can be computed as
\begin{equation}
\label{eq_center_offset}
    y_{\text{width}}^{t,c}=\frac{e-b}{R}\, , \quad y_{\text{offset}}^{t,c}=\frac{b+e}{2S}-c_t \, .
\end{equation}
The localization loss is then simply the L1 distance between ground-truth and predicted values.

\subsection{Inference}

At inference, the model outputs a sequence of six prediction steps every 240ms, which accounts almost entirely for the user perceived latency. At step $t$, a score is computed as $\max\{\hat{y}_{\text{class}}^{t,c}:c\leq C\}$. If the score is above a threshold $\vartheta$, then an event is proposed as $\big(\argmax\{\hat{y}_{\text{class}}^{t,c}:c\leq C\}, \hat{b}_t, \hat{e}_t\big)$ where $\hat{b}_t$ and $\hat{e}_t$ are the estimated begin and end timings of the event computed from the width and offset predictions. Their relation is defined in Equation \eqref{eq_center_offset}. Non-maximum suppression (NMS) \cite{Neubeck2006} is then used to select the best non-overlapping proposals based on their scores and timings, which suppresses repetitive proposals.

%%%% EXPERIMENTS %%%%
\section{Experiments}

%%% Table 1 %%%%
\begin{table*}[ht!]
\centering
\caption{Metrics comparison between our models and a BC-ResNet baseline that was reproduced from \cite{Samragh2023}. Bold values indicate that a metric improves upon BC-ResNet. The last column shows the portion of skipped MACs when using gates.}
\label{table:libri}
\resizebox{1.0\textwidth}{!}{
\begin{tabular}{l l c c c c c c c c c c} 
\toprule
\textbf{Data} & \textbf{Model} &  \textbf{Params} & \textbf{FRR} & \textbf{FAR} & \textbf{Precision} & \textbf{Recall} & \textbf{F1} & \textbf{Actual} & \textbf{IOU} & \textbf{MTWV} & \textbf{Skips $[\%]$} \\ [0.5ex] 
\midrule
Librispeech & BC-ResNet-L & 1.54M & 0.132 & 0.192 & 0.856 & 0.868 & 0.862 & 0.872 & 0.850 & 0.78 & 0\\
\textit{test-clean} & Ours-L    & 1.29M & \textbf{0.129} & \textbf{0.097} & \textbf{0.922} & \textbf{0.871} & \textbf{0.896} & 0.861 & 0.839 & \textbf{0.87} & 0\\
 & Ours-L-gated & 1.29M & 0.134 & \textbf{0.100} & \textbf{0.920} & 0.866 & \textbf{0.892} & 0.866 & 0.830 & \textbf{0.84} & 30\\
& Ours-S    & 966k & 0.168 & \textbf{0.107} & \textbf{0.911} & 0.832 & \textbf{0.870} & 0.821 & 0.839 & \textbf{0.84} & 0\\
%Ours-S-gated & 975k & 0.172 & \textbf{0.107} & \textbf{0.910} & 0.828 & \textbf{0.867} & 0.823 & 0.837 & \textbf{0.83} & 20$\%$\\
\bottomrule \\[-1.5ex]
Librispeech & BC-ResNet-L & 1.54M & 0.316 & 0.314 & 0.749 & 0.684 & 0.715 & 0.684 & 0.843 & 0.58 & 0\\
\textit{test-other} & Ours-L & 1.29M & \textbf{0.295} & \textbf{0.146} & \textbf{0.869} & \textbf{0.705} & \textbf{0.779} & \textbf{0.697} & 0.830 & \textbf{0.76} & 0\\
 & Ours-L-gated & 1.29M & \textbf{0.306} & \textbf{0.151} & \textbf{0.863} & \textbf{0.694} & \textbf{0.769} & \textbf{0.688} & 0.820 & \textbf{0.70} & 26\\
& Ours-S    & 966k & 0.354 & \textbf{0.145} & \textbf{0.859} & 0.646 & \textbf{0.737} & 0.646 & 0.830 & \textbf{0.67} & 0\\
%Ours-S-gated & 975k & 0.356 & \textbf{0.148} & \textbf{0.856} & 0.644 & \textbf{0.735} & 0.629 & 0.827 & \textbf{0.68} & 26$\%$\\
\bottomrule \\[-1.5ex]
GSC & BC-ResNet-XS & 139k & 0.055 & 0.004 & 0.966 & 0.945 & 0.955 & 0.945 & 0.762 & 0.84 & 0\\
 & Ours-XS & 93k & \textbf{0.052} & \textbf{0.002} & \textbf{0.982} & \textbf{0.948} & \textbf{0.964} & \textbf{0.948} & \textbf{0.818} & \textbf{0.89} & 0\\
 & Ours-XS-gated & 94k & 0.056 & \textbf{0.003} & \textbf{0.976} & 0.944 & \textbf{0.960} & 0.944 & 0.757 &  \textbf{0.87} & 42, 97\\
\bottomrule
\end{tabular}
}
\end{table*}

%%%% Libsrispeech %%%%
\subsection{Top1000 Librispeech}
\label{libri}

%%%% Dataset %%%%
\textbf{Dataset.} 
The Librispeech dataset \cite{librispeech} is an English corpus obtained from audio books that have been read aloud and sampled at a frequency of 16 kHz. The training set contains 280k utterances, totaling 960 hours of speech. It is used for keyword spotting by defining a lexicon with the 1000 words that appear the most frequently within the training set. Similarly to \cite{Segal2019, Samragh2023}, the start and end timings of the words are extracted using the Montreal Forced Aligner \cite{Mcauliffe2017}. It represents a rather difficult KWS task as (i) many keywords often collide inside a single window due to the continuous read speech nature of the utterances, and (ii) many confusable words such as (peace, piece), (night, knight), and (right, write) are treated as distinct classes. The results are reported on \textit{test-clean} and \textit{test-other}, where the latter represents more challenging data. \\[+0.5ex]
%%%% Training %%%%
\noindent\textbf{Training.} 
Gated and non-gated conformer architectures with $H=80$ and $N_z=8$ are trained with the Adam optimizer \cite{Kingma2015} and a Cosine Annealing scheduler, which gradually reduces the learning rate from 0.001 to 0.0001 over 100 epochs. Data augmentation is applied by randomly cropping utterances before the start of the first keyword, which makes the model agnostic to shifts. Additionally, zero-padding both sides with 250ms ensures that no audio portion gets discarded during the windowing procedure. All models are trained on eight NVIDIA-V100 GPUs using Pytorch distributed data parallelism \cite{Li2020} and a batch-size of eight utterances per GPU. \\[+0.5ex]
%%%% Evaluation %%%%
\noindent\textbf{Evaluation.} 
To evaluate a trained model, we first obtain predicted events with NMS score greater than $\vartheta=0.95$ and count as true positives (TPs) the ones for which a ground-truth event of the same class overlaps. Ground-truth events that are not predicted by the model are counted as false negatives (FNs), and predicted events that do not have a corresponding ground-truth of the same class are counted as false positives (FPs). This allows us to compute precision, recall, F1-score, actual accuracy and average IOU similarly to \cite{Segal2019, Samragh2023}. Mean term weight values (MTWV) \cite{Fiscus2007} are also reported using twenty selected keywords originally chosen in \cite{Palaz2016} and used in \cite{Segal2019, Samragh2023}, where $\vartheta$ is tuned for each keyword. False reject rate (FRR) is computed as $\text{FNs} / (\text{FNs}+\text{TPs})$ and false accept rate (FAR) as FPs per second. The portion of skipped multiply-and-accumulate operations (MACs) in the conformer over the complete test set is also reported to illustrate the efficiency benefits of using gates. \\[+0.5ex]
%%%% Results %%%%
\noindent\textbf{Results.} 
As presented in the top and middle panels of Table \ref{table:libri}, while using 16$\%$ fewer parameters, our approach improves upon the modified BC-ResNet baseline defined in \cite{Samragh2023} on almost all metrics, especially on \textit{test-other}. 
%In particular, the FAR gets halved, hence the significant increase in F1-score and MTWV. %The lower IOU can be partly explained by the larger receptive field of our models with 1s compared to 825ms for the BC-ResNet. 
Adding input-dependent dynamic gates to the encoder (Ours-L-gated) results in skipping on average 30$\%$ and 26$\%$ of MACs on \textit{test-clean} and \textit{test-other} respectively while maintaining the same performance (less than $1\%$ difference on all metrics). It also outperforms a non-gated smaller model (Ours-S) with comparable number of MACs, hence the benefits of the I3D method.

%%%% GSC %%%%
\subsection{Google speech commands}

%%%% Dataset %%%%
\noindent\textbf{Dataset.}
We place the 35 Google speech commands v2 \cite{Warden2018} in isolation over a stream of babble noises from the MS-SNSD dataset \cite{Reddy2019} with signal to noise ratio of 10-40dB. \\[+0.5ex]
%%%% Training %%%%
\noindent\textbf{Training.} 
Here tiny conformer architectures with $H=40$ and $N_z=3$ are trained as explained in \ref{libri}. For the BC-ResNet baseline, we use the same architecture as on Librispeech with a reduced number of convolution channels.
%%%% Results %%%%
\noindent\textbf{Evaluation.} 
The evaluation is similar to that on Librispeech except here all labels are used to compute MTWVs, and skipped MACs are reported for speech and non-speech inputs separately. \\
%%%% Results %%%%
\noindent\textbf{Results.} 
Although our approach still improves upon the BC-ResNet baseline with 32$\%$ fewer parameters, this simpler task mainly aims to demonstrate the benefits of gates in a command scenario. Here the encoder shows its ability to distinguish between speech and non-speech inputs and adapt its processing accordingly. We indeed measure that 42$\%$ of MACs are skipped when processing regions containing speech, compared to 97$\%$ for pure noise. As expected, the computational savings are even more significant in the absence of commands, making our method particularly interesting to improve the efficiency of always-on models.

%%%% Figure 2 %%%%
%\begin{figure}[h]
%\centering
%\includegraphics[width=1.0\columnwidth]{images/det_curves.png}
%\caption{Det curve comparison with baseline.}
%\label{fig:detcurve} 
%\end{figure}

%%%% CONLUSION %%%%
\section{Conclusion}

This paper proposes a method for efficient and streaming KWS. We incorporate a streaming conformer encoder into a vision-inspired KWS pipeline and include trainable binary gates to control the network's dynamic depth. These gates can selectively skip modules based on input audio characteristics, resulting in reduced computations. 
Our method outperforms the baseline in both continuous speech and isolated command tasks, while using fewer parameters, thereby maintaining a small memory footprint. Furthermore, the gates allow us to considerably reduce the average number of computations during inference without affecting the overall performance. Their inclusion is observed to be even more advantageous in a scenario where speech commands appear sparsely over some background noise.

%%%% REFERENCES %%%%
%\bibliographystyle{IEEEbib.bst}
%\bibliography{refs}
\vfill\pagebreak
\section{References}
\ninept
\printbibliography[heading=none]

\end{document}